# Structural Flexibility of the TCF7L2–DNA Complex with the Type 2 Diabetes SNP rs7903146

Karthik Venuturimilli[1*\[0009-0002-3664-2155\]], Yang Ha[1\[0000-0001-5684-8420\]]

[1]Lawrence Berkeley National Lab,
Berkeley CA, United States
`karthikv@lbl.gov`

**Abstract.** The single nucleotide polymorphism (SNP) rs7903146 in the TCF7L2 gene has been determined as one of the strongest common genetic risk factors for Type 2 Diabetes (T2D). The location of the SNP in a non-coding region suggests a regulatory mechanism, meaning the SNP doesn't change the protein's own structure but rather affects how the TCF7L2 protein binds to DNA to control other genes. This binding, however, is highly dependent on the shape and flexibility of the DNA. This study aims to reveal the atomic-level effects of the SNP's cytosine-to-thymine substitution on the TCF7L2-DNA complex. We first utilized AlphaFold to generate individual high-confidence structures of the TCF7L2 protein and two 15-base pair DNA duplexes: one containing the reference C allele and one containing the variant T allele. These structures were then used as inputs for Neurosnap's Boltz2 deep learning model to generate two complete protein-DNA complexes of the TCF7L2 HMG-box bound to each DNA variant. Using the iMODS server, we conducted a Normal Mode Analysis (NMA) to predict and compare large-scale flexibility and differences in interactions between the complexes. The protein-DNA interface was dissected using PDBsum to locate atomic contacts, clefts, and interaction maps. Overall, our results show that the T allele variant exhibits increased global stiffness with a higher eigenvalue and reduced flexibility, suggesting that the SNP disrupts the mechanism and biomechanical balance needed for efficient TCF7L2-DNA binding, thus affecting downstream gene regulation.

## 1. Introduction

Type 2 Diabetes (T2D) is among the most common metabolic diseases globally, affecting hundreds of millions of people and causing enormous global health and economic burdens. Although external elements like lifestyle and nutrition play major roles, genetic predisposition is a significant determinant of T2D. Of all the known loci, the Transcription Factor 7-Like 2 (TCF7L2) gene has been identified as the strongest common genetic risk factor for T2D in genome-wide association studies (GWAS) [1], [2], [3].

The rs7903146 single-nucleotide polymorphism (SNP) in the TCF7L2 gene, with a substitution of cytosine-to-thymine (C→T), has emerged as a strongly associated polymorphism with impaired insulin secretion and glucose homeostasis [1], [2], [4], [5]. This SNP is located in a non-coding intron, indicating that the effect is most likely to be regulatory instead of structural. This, in turn, means that the polymorphism perhaps does not alter the amino acid sequence of TCF7L2 itself, but could affect how the TCF7L2 protein binds to DNA, affecting downstream gene transcription within the Wnt signaling pathway that modulates β-cell function and glucose metabolism [4], [5], [6].

Recent studies have shown that TCF7L2 binding is cell-type specific and can occur by both direct DNA binding as well as cooperative tethering to other transcription factors, such as GATA3 and FOXA2 [7]. These context-specific modes of TCF7L2 binding illustrate the complexity of the TCF7L2-mediated transcriptional regulation, where even tiny alterations to DNA conformation or binding energetics can cause altered gene expression profiles.

Protein-DNA recognition driven by transcription factors like TCF7L2 depends highly upon DNA flexibility and shape, in combination with the protein's High Mobility Group (HMG) box domain's conformational flexibility [8], [9]. The HMG box is a DNA-binding motif that can cause the duplex to distort and bend in an effort to stabilize nucleoprotein complexes. Studies on these structures have shown that HMG-box proteins cause extensive DNA curvature and change local groove geometry upon DNA binding [8]. These types of deformations can modulate the binding specificity and can contribute significantly in transcriptional regulation. It is important to understand these subtle effects at an atomic level in order to define the mechanism by which rs7903146 plays a role in the pathogenesis of T2D.

In this study, we investigate the biophysical effects of the rs7903146 SNP on the TCF7L2–DNA complex with a combination of advanced computational and AI tools. High-confidence protein–DNA models of the reference (C) and variant (T) alleles were prepared using Google DeepMind's AlphaFold AI model [10] and Boltz2 deep-learning models by Neurosnap [11], [12], [13]. The resulting complexes were explored through Normal Mode Analysis (NMA) using iMODS [14] to investigate large-scale flexibility and collective motions and PDBsum [15] to visualize atomic contacts and interfacial hydrogen bonds. Ultimately, by bringing these procedures together, we seek to offer an atomic-level explanation of how the substitution at rs7903146 disrupts the mechanics of the TCF7L2–DNA interface, giving structure-based insight into its contribution to Type 2 Diabetes susceptibility.

## 2. Methods

### 2.1 Protein Structure Prediction Using AlphaFold

The 3D structure of the TCF7L2 protein was generated using Google DeepMind's latest model, AlphaFold3 [10]. We retrieved the full-length sequence of the protein from the UniProt database. The sequence was then truncated to reflect only the functional HMG-box domain. This sequence was submitted to the AlphaFold web server, and the top-ranked model was selected based on a predicted Local Distance Difference Test (plDDT) score >90 across the entire structure, indicating high prediction confidence. This structure was then validated in PyMOL v3.1 [16] (Schrödinger, LLC) to confirm the proper folding of the HMG-box domain.

### 2.2 DNA Structure Prediction and Protein-DNA Complex Formation

To analyze the effects of the rs7903146 SNP on structure and flexibility, we modeled a 15-base pair DNA double helix surrounding the polymorphic site for both the reference with the C allele and the variant with the T allele. The exact flanking sequences surrounding the SNP were found using the National Center for Biotechnology Information's (NCBI) dbSNP database [17]. With seven base pairs flanking each side of the SNP, the sequences were input into AlphaFold [10] to generate high plDDT double-stranded structures for both the reference C allele and variant T allele structures. The Protein-DNA complex assembly was created using Neurosnap's Boltz2 deep learning model [11], [12], [13]. This platform predicts optimal protein-nucleic acid configurations based on AlphaFold databases and learned geometric and electrostatic features. The resulting complexes generated by Neurosnap were presented through the same confidence scoring mechanism as AlphaFold, with high plDDT scores of >90 for both complexes' structures. We then examined these complexes to ensure minor groove insertion of the TCF7L2 HMG-box using PyMOL [16].

**2.3 Normal Mode Analysis using iMODS**

Normal Mode Analysis (NMA) was conducted to assess the large-scale conformational dynamics and flexibility of both complexes using the iMODS web server [14]. The server uses an internal coordinate elastic network model to compute the eigenvalues, which are scalar values that represent the energy required to deform the structure along a motion. Other outputs we analyzed from iMODS include deformability, B-factors that describe how much each atom and residue fluctuates around its average position, and covariance maps that describe the overall motion of the macromolecules. For both models, the lowest-frequency normal mode, labeled as mode 1 in iMODS, was analyzed to describe the collective motions of the complexes. The eigenvalue gave insight into structural stiffness, and the deformability and covariance plots and B-Factor profiles were automatically generated and given by iMODS outputs. We organized all numerical values for comparative analysis between the two structures.

**2.4 Interface Analysis by PDBsum**

Additional detailed protein-DNA interaction assessments and analyses were performed using the PDBsum web server [15], which automatically identifies and visualizes interface residues and classifies them into hydrogen bonds, non-bonded contacts, and cleft geometries for any provided custom PDB structure[18]. For both complexes, PDBsum generated bond contact maps that visually showed types of bonds between phosphates/nucleotides and the protein's specific amino acids. PDBsum specifically identified hydrogen bonds and non-bonded contacts. Cleft and surface analysis was also done, reporting cleft volume and buried vertices. We used the cleft volume changes and differences in key residues based on the contact maps to infer changes in interface compactness and binding flexibility between the two complexes. To generate detailed figures, structural visualizations and alignments were done using PyMOL [16], while normal mode animations from iMODS [14], and the interaction maps from PDBsum [15] were analyzed and presented.

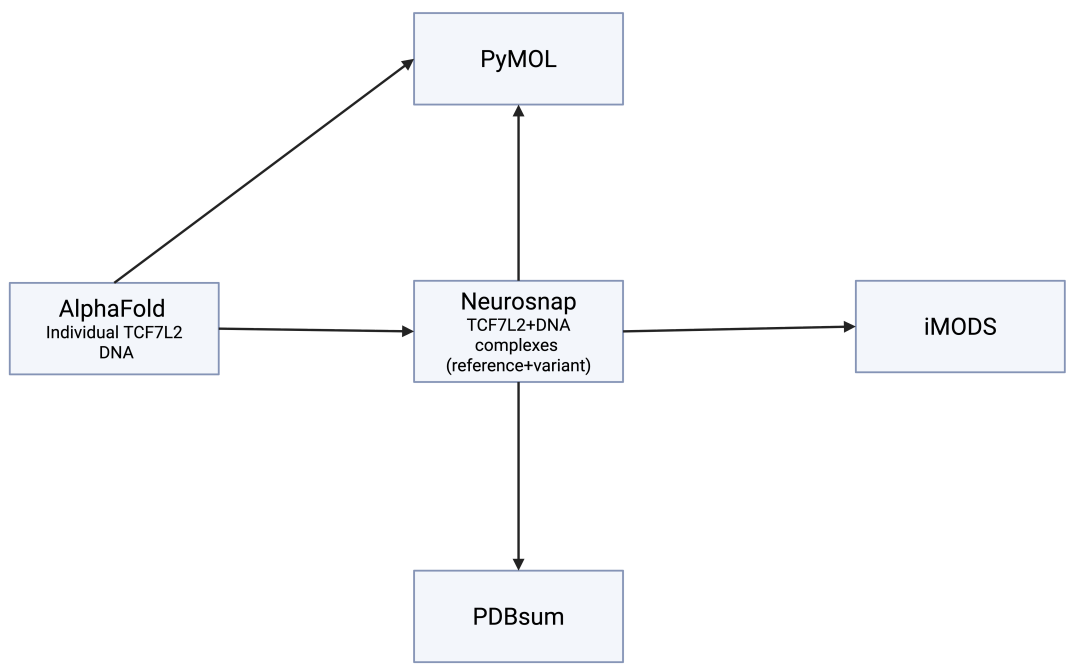

Fig. 1. Computational workflow for modeling and analysis of the TCF7L2–DNA complex. Created in BioRender. Venuturimilli, K. (2025) https://BioRender.com/125ivz1 [19]

## 3. Results

**3.1 Normal Mode Analysis Reveals Heightened Structural Rigidity in Variant Complex**

iMODS Normal Mode Analysis (NMA) [14] outputted clear distinctions in the flexibility patterns between the reference and variant TCF7L2-DNA complexes. The reference complex showed a smaller eigenvalue ($5.35 \times 10^{-4}$) compared to the variant ($8.36 \times 10^{-4}$), illustrating that the reference conformation needed less energy for deformation and was consequently more flexible (Table 1). Overall, the higher eigenvalue exhibited by the variant indicates increased stiffness within the complex and repressed large-scale motions.

Analyzing the B-factor (mobility) profiles further validated the discrepancies we found through the eigenvalues. The reference model exhibited various peaks throughout the B-Factor plot, which indicates that more flexible regions are present in the DNA-binding interface and HMG-box domain. However, the variant, though also exhibiting peaks, presented a less exaggerated profile, which suggested reduced residue-level motion. The peak numbers that exhibited high flexibility ($> 0.6$) declined from five in the reference to one in the variant, which signifies a loss of mobility in localized regions of the interface..

Visualizations of the most prominent normal mode (Mode 1) showed that the motion was qualitatively different as well. The reference complex showed an outward hinge-like bending motion where the bottom DNA arm swung away from the protein, in agreement with dynamic DNA bending. Contrastingly, the variant complex demonstrated an inward motion that was torsionally compact, where both the top and bottom parts of the DNA complex, above the areas of binding between the DNA and protein, bent inwards on itself, illustrating a more compact structure, consistent with our quantitative findings.

Table 1, Fig. 2 and Fig 3 show the mobility and deformability profiles of the two complexes, as well as the Mode 1 visualizations. Overall, these values and visualizations indicate that the single T allele substitution induced mechanical stiffness and reduced the flexibility of the complex as a whole, reducing the efficiency for DNA binding. The variance plots show how much of the total possible motion is captured by each mode, and the variant's variance plots require more modes to reach a level of 90% cumulative motion, compared to the reference plots, signifying constrained dynamic behavior.

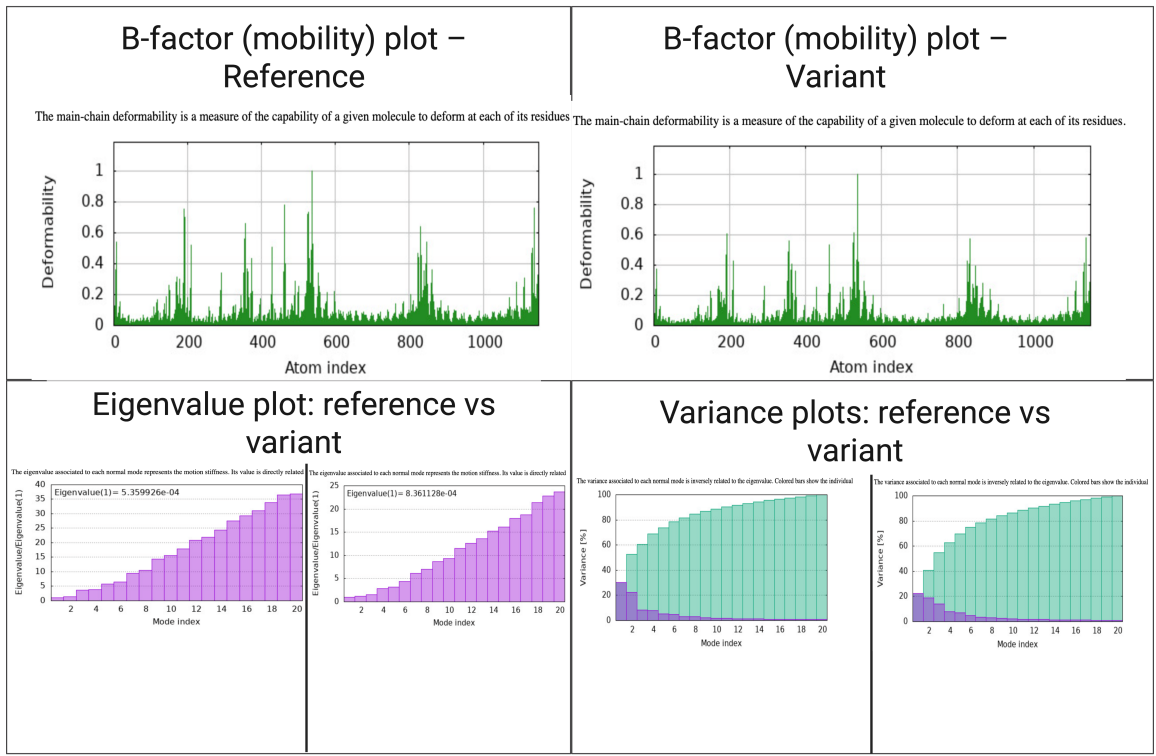

**Fig. 2**. Normal Mode Analysis Results Comparing the Reference and Variant TCF7L2–DNA Created in BioRender. Venuturimilli, K. (2025) https://BioRender.com/voa0whc [19]

**Table 1**. Normal Mode Analysis (iMODS) Results for the Reference and Variant TCF7L2–DNA Complexes

| Parameter | Reference (C allele) | Variant (T allele) | Interpretation/ Notes |
|---|---|---|---|
| Eigenvalue (Mode 1) | $5.35 \times 10^{-4}$ | $8.36 \times 10^{-4}$ | |
| Mean deformability | $4.95 \times 10^{-3}$ | $5.82 \times 10^{-3}$ | Slightly higher local deformability, but overall restricted motion |
| Min deformability | $6.95 \times 10^{-4}$ | $8.24 \times 10^{-4}$ | Variant has more localized spikes in flexibility, but fewer distributed peaks |
| Max deformability | $6.34 \times 10^{-2}$ | $9.36 \times 10^{-2}$ | |
| High-flexibility peaks (>0.8) | 1 | 1 | |
| Flexibility peaks (>0.6) | 5 | 1 | Reference complex has broader flexible interface regions. |
| Cumulative variance (first 5 modes) | 65% | 57% | Variant complex requires more modes to represent total motion |
| Cumulative variance (first 10 modes) | 90% | 85% | |
| Cumulative variance (20 modes) | 99% | 99% | Both show similar total dynamic range |

| Covariance map | Uniformly red | Uniformly red | |
|---|---|---|---|
| Mode 1 motion | Lower DNA bends away; HMG-box flexes outward | Both DNA arms bend inward; torsionally compact motion | Overall variant shows less adaptive bending - more constrained |

Reference vs Variant Mode 1 end of animation complexes

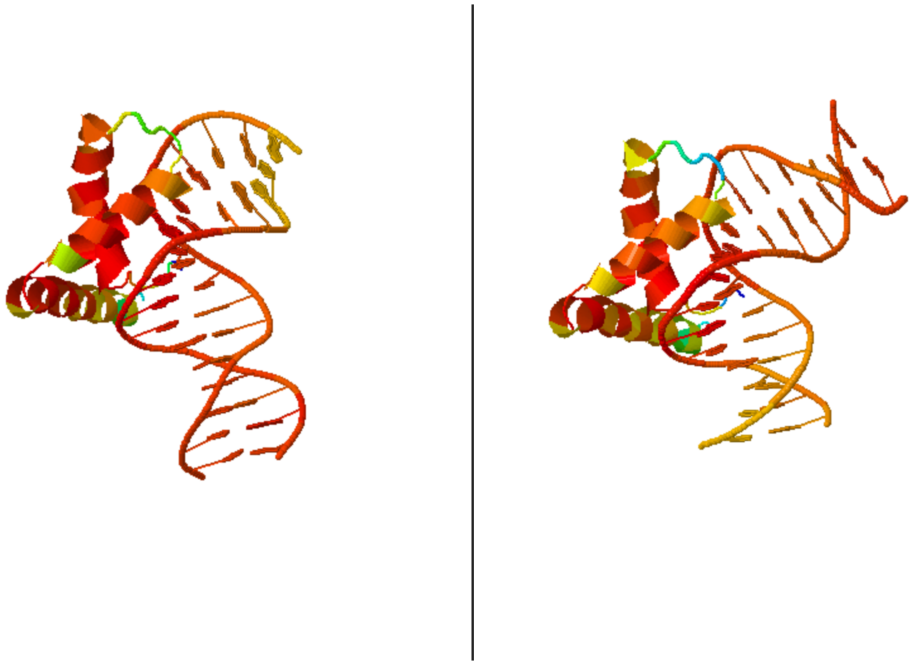

**Fig. 3**. iMODS Reference and Variant Model 1 animation TCF7L2–DNA Complexes. Created in BioRender. Venuturimilli, K. (2025) https://BioRender.com/1nqahkh [19]

## 3.2 Interface Mapping Highlights Altered Contact Networks in the Variant Complex

The protein-DNA interface analysis conducted using PDBsum [15] has revealed quantitative changes in the interaction networks between the complexes. The reference complex formed five hydrogen bonds and 29 non-bonded contacts, with a cleft volume of 3142.55 Å³ and six buried vertices (Table 2). These values indicate an open, stable interface. The large cleft volume illustrates a much broader and extended interface geometry. This allows the HMG box to flex around the DNA backbone, also allowing the DNA to move more freely within the space. It also allows the box to stabilize through multiple distributed interactions.

The variant complex, on the other hand, also had five hydrogen bonds, but only 21 non-bonded contacts and a reduced cleft volume of 1763.02 Å³ (Table 2), demonstrating a much tighter, compact binding region. These results of contraction, reflected by the cleft volume and bonding identities, mirror the increased rigidity observed in our Normal Mode Analysis [14].

PDBsum's [15] residue mapping, shown in Figures 4 and 5, also depicts multiple changes in key residue bindings. The reference complex relied on interactions involving Lys2, Arg16, Ser28, Asn32, and Tyr51, with multiple phosphate-mediated stabilizations by Arg55 and Arg58 along the backbone (Table 2). The variant complex preserved interactions with Lys2, Arg16, Ser28, Tyr51, and Asn31, but actually gained new interactions with Glu27 and Phe8, while losing Arg55 and Arg58, residues that had been anchoring the complex at key phosphate positions for the reference complex (Table 2).

Overall, these distinctions may stabilize the variant complex locally through the new polar contacts and bindings, but reduce flexibility and the diversity of contacts, limiting the complex's ability to adapt to conformational changes during binding.

The PDBsum cleft diagrams (Fig.6) show these differences with cleft pocket visualizations for both complexes.

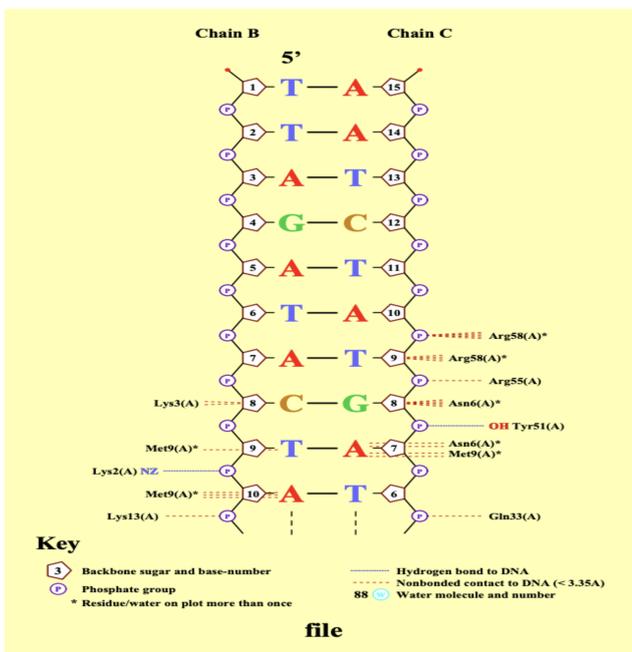

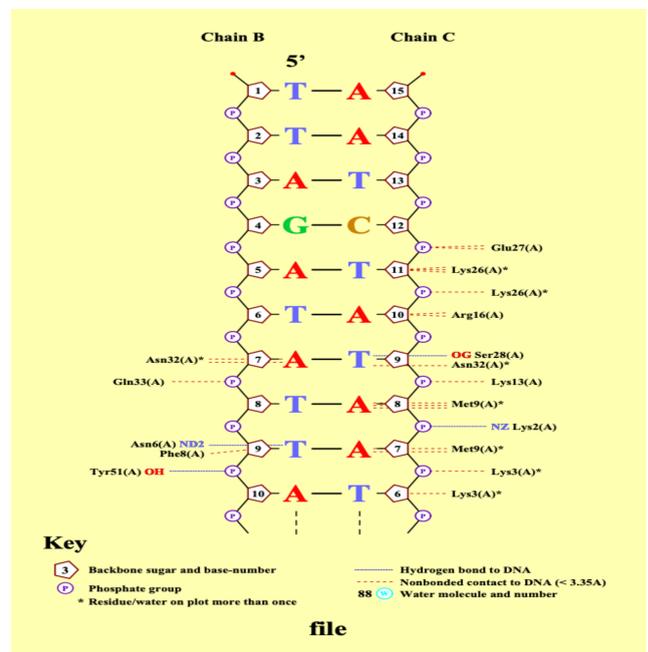

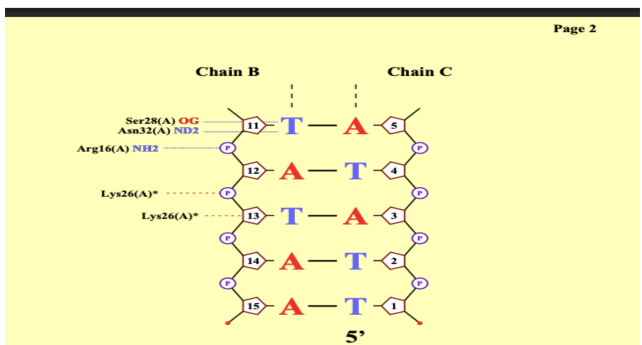

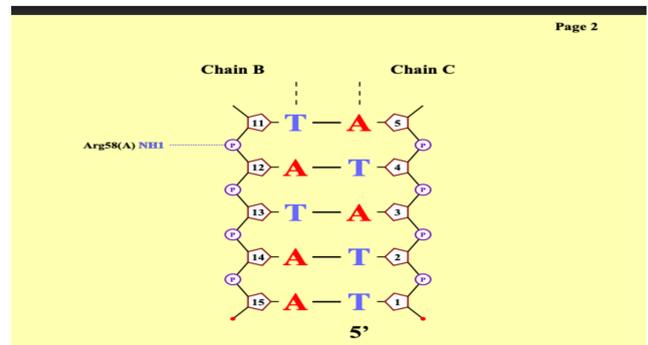

**Fig. 4**. Reference Residue Mapping by PDB sum of TCF7L2–DNA Complex. Created in BioRender. Venuturimilli, K. (2025) https://BioRender.com/13l73jf [19]

**Fig. 5**. Variant Residue Mapping by PDB sum of TCF7L2–DNA Complex. Created in BioRender. Venuturimilli, K. (2025) https://BioRender.com/88584ju [19]

**Table 2**. Protein-DNA Interface Parameters from PDBsum Analysis

| Parameter | Reference (C allele) | Variant (T allele) | Interpretation/Notes |
|---|---|---|---|
| Hydrogen Bonds | 5 | 5 | Hydrogen Bonding Network conserved, indicating stability locally for both complexes |
| Non-Bonded contacts | 29 | 21 | Reduced interface diversity for the variant due to the loss of side chain DNA interactions in the variant |
| Cleft Volume (Å$^3$) | 3142.55 | 1762.02 | Largely reduced cleft volume indicates a much more compact and less accessible variant interface |
| Buried Vertices | 6 | 7 ( rounded from 6.63) | |
| Key Residues | Lys2, Arg16, Ser28, Asn32, Tyr51, Arg55, Arg58, Asn6, Met9, Lys13, Gln33, Lys26 | Lys2, Arg16, Ser28, Asn32, Tyr51, Glu27, Phe8, Lys26, Arg58 (weakened), Met9, Lys3, Gln33 | Multiple phosphate mediated interactions lost, (Arg55 and Arg58) leading to reduced flexibility overall |

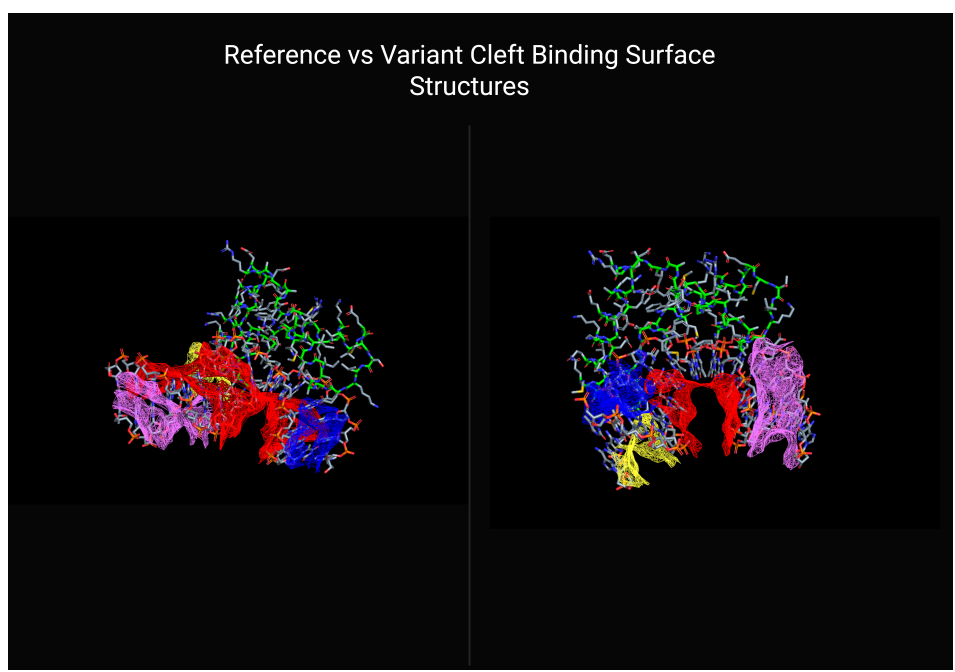

**Fig. 6**. Reference vs Variant Cleft Binding Surface Structures of TCF7L2–DNA

## 4. Conclusion

This study computationally examined the structural effects of the SNP rs7903146, with its cytosine to thymine substitution within the TCF7L2-DNA complex using iMODS, PDBsum, and PyMOL [14], [15], [16]. Normal Mode Analysis [14] revealed that the variant complex exhibited a higher overall eigenvalue ($8.36 \times 10^{-4}$) compared to the reference ($5.35 \times 10^{-4}$), indicating increased rigidity and reduced overall flexibility within the complex. PDBsum [15] further validated these results by showing a significantly reduced cleft volume in the variant complex (1763.02 Å³) compared to the reference (3142.55 Å³), consistent with a more compact and less adaptable protein-DNA interface. The B-factor profiles revealed that high-flexibility peaks (>0.6) declined from five in the reference to only one in the variant, showing us a loss of mobility in localized regions of the interface.

Based on these findings, we conclude that the T allele variant exhibits increased global stiffness with a higher eigenvalue and reduced flexibility. The rs7903146 SNP disrupts the biomechanical balance needed for efficient TCF7L2-DNA binding through three primary mechanisms: (1) increased structural rigidity that restricts large-scale conformational motions, (2) reduced cleft volume that limits DNA flexibility within the binding interface, and (3) altered residue contact networks that decrease the diversity of stabilizing interactions. The loss of specific critical phosphate-mediated interactions by Arg55 and Arg58 in the variant complex reduces the protein's ability to maintain optimal DNA binding geometry.

Overall, these structural changes impair efficient DNA recognition and transcriptional regulation, which may disrupt TCF7L2's role in regulating genes involved in glucose metabolism and insulin secretion within the Wnt signaling pathway, thus affecting downstream gene regulation and increasing the risk of developing Type 2 Diabetes. The T allele substitution alters the mechanical dynamics of the TCF7L2-DNA interface at the atomic level, providing a molecular explanation for the strong association between rs7903146 and T2D susceptibility observed in genome-wide association studies. Future work will include molecular dynamics simulations and binding free energy analyses to quantify stability changes.